\documentclass[twoside,fleqn]{article}
\usepackage{espcrc2}

\usepackage{graphicx}
\makeatletter
\advance \parskip by 0pt plus 0.1pt minus 0.1pt\relax
\newcommand\ds{\displaystyle}
\newcommand\etal{{\it et al.\spacefactor1000}}
\newcommand\ibid{{\it ibid.\spacefactor1000}}

\newcommand\nameaddress[1]{{\addtocounter{address}\m@ne
                            \expandafter\xdef
                               \csname address.#1\endcsname
                               {\number\value{address}}%
                            \addtocounter{address}\@ne}}
\newcommand\useaddress[1]{{\edef\doit{\noexpand\addressmark
                                      \noexpand\setcounter{address}%
                                      {\number\value{address}}}%
                           \setcounter{address}{\csname address.#1\endcsname
                                                }%
                           \expandafter}\doit}
\newcommand\anotheraddress[1]{{\let\orig@makeadmark=\@makeadmark
                               \def\@makeadmark##1{\orig@makeadmark
                                                   {\negthinspace,##1}}%
                               \address{#1}}}
\newcommand\slashnext[1]{\mathpalette{\bgroup\let\style=}
                                     {\setbox0=\hbox{$\style #1$}%
                                      \setbox2=\hbox to\wd0{\hss$\style/$\hss}%
                                      \hbox to 0pt{\box2\hss}\box0\egroup}}
\@ifundefined{inlinecite}{\let\inlinecite=\cite}
\makeatother

\title{\begin{flushright}\normalsize
       \end{flushright}
	Order $a$ improved renormalization constants}

\author{T. Bhattacharya,\address{MS B285, Los Alamos National Lab, Los Alamos,
                 New Mexico 87545, USA}\nameaddress{lanl} 
	R. Gupta,\useaddress{lanl}
	W. Lee,\useaddress{lanl}
        and
	S. Sharpe\address{Physics Department, University of Washington,
		          Seattle, Washington 98195, USA}%
        }
       
\begin{document}

\begin{abstract}
We impose the axial and vector Ward identities on local fermion
bilinear operators in the Sheikholesalami-Wohlert discretization of
fermions.  From this we obtain all the coefficients needed to improve
the theory at \(O(a)\), as well as the scale and scheme independent
renormalization constants, \(Z_A\), \(Z_V\) and \(Z_S/Z_P\).  
\end{abstract}

\maketitle

\section{Introduction}
\vspace{-3.6pt}
In Reference \inlinecite{lanl98}, we studied the improvement of the
Wilson-Dirac theory by removing all lattice artifacts linear in the
lattice spacing \(a\) in on-shell matrix elements.  In general, such
an improvement for the dimension 3 fermion bilinear operators requires
one to tune the coefficient of the clover operator in the action,
determine the coefficient of an extra term each in the vector,
axial-vector, and tensor currents, and obtain the mass dependence of
all the renormalization constants.  We showed that all these
improvement constants, as well as the renormalization constants for
the axial and vector currents, can be determined by imposing the axial
and vector Ward identities.  The remaining renormalization constants,
those for the scalar, pseudoscalar and tensor operators, are scheme
dependent and cannot be determined by this method.  The scheme
dependence is, however, the same for both the scalar and the
pseudoscalar operators, and the corresponding ratio of renormalization
constants can again be obtained.

To test the efficacy of this method, we had evaluated the
renormalization constants at \(\beta=6.0\) for the perturbative
(tadpole improved) value of the clover coefficient, and found that
these differed from the values determined by the ALPHA
collaboration~\cite{alpha} using a non-perturbatively tuned
coefficient.  It was not clear whether the differences arose from
different \(O(a^2)\) errors in the two calculation, or whether the
difference in the clover coefficient was responsible for the variation
observed.  Here we report results of carrying out our procedure after
changing the clover coefficient to that used by the ALPHA
collaboration, and also at \(\beta=6.2\), where the \(O(a^2)\) errors
are expected to be much smaller.

For the details of the notation used, the numerical technique, a study
of the consistency when the same constant is determined in multiple
ways, and the choice of the method which determines each quantity
best, we refer the reader to Reference~\inlinecite{lanl98}.  To avoid
confusion, we repeat here that the coefficients \(\tilde b_X\) we
defined differ from the coefficients \(b_X\) used by earlier
authors.  In particular, at the level of \(O(a)\) improvement, one
has \( \tilde b_X = (Z_A^0 Z_S^0 / Z_P^0)  b_X \).

\section{Lattice parameters and Results}
\begin{table*}
\caption{Renormalization constants and improvement coefficients at
         \(\beta=6.0\) and \(\beta=6.2\).}
\label{tb:1}
\setlength{\tabcolsep}{0.5pt}
\begin{tabular}{||c||l|l|l|l||l|l|l||}
\hline
\multicolumn{1}{||c||}{}&
\multicolumn{4}{c||}{\(\beta=6.0\)}&
\multicolumn{3}{c||}{\(\beta=6.2\)}\\
\hline
         & LANL              & LANL             & ALPHA        & P. T.
         &  LANL             &  ALPHA           &  Pert. Th.     \\
         &                   &                  &              &
         &                   &                  &                \\[-12pt]
\hline			     		       
         &                   &                  &              &
         &                   &                  &                \\[-12pt]
$C_{SW}$ & 1.4755            & 1.769            & 1.769        &  1.4785
         &  1.614            &  1.614           &  1.442         \\
         &                   &                  &              &
         &                   &                  &                \\[-12pt]
\hline			     		       
         &                   &                  &              &
         &                   &                  &                \\[-12pt]
$Z^0_V$  & $+0.745(1)(1)$    & $+0.764(1)(1)$   & $0.7809(6)$  & $+0.810(5)$  
         &  $+0.784(00)(00)$ &  $+0.7922(4)(9)$ &  $+0.821(5)$   \\
$Z^0_A$  & $+0.76 (2)(1)$    & $+0.811(9)(12)$  & $0.7906(94)$ & $+0.829(3)$  
         &  $+0.816(08)(05)$ &  $+0.807(8)(2) $ &  $+0.839(2)$   \\
$Z^0_P/Z^0_S$		     		       				    
         & $+0.77 (4)(1)$    & $+0.86 (2)(0)$   &  N.A.        & $+0.949(3)$  
         &  $+0.878(15)(06)$ &  N.A.            & $+0.953(2)$    \\
         &                   &                  &              &              
         &                   &                  &                \\[-12pt]
\hline			     		       
         &                   &                  &              &              
         &                   &                  &                \\[-12pt]
$c_A$    & $-0.005(15)(20)$  & $-0.02(2)(2)$    & $-0.083(5)$  & $-0.013(0)$  
         &  $-0.033(04)(03)$ &  $-0.038(04)$    &  $-0.012(1)$   \\
$c_V$    & $-0.66 (27)(02)$  & $-0.02(15)(07)$  & $-0.32 (7)$  & $-0.028(1)$  
         &  $-0.10 (14)(06)$ &  $-0.21(7)$      &  $-0.026(7)$   \\
$c_T$    & $+0.17 (07)(01)$  & $+0.05(04)(00)$  &  N.A.        & $+0.020(0)$  
         &  $+0.062(35)(08)$ &  N.A.            &  $+0.019(4)$   \\
         &                   &                  &              &              
         &                   &                  &                \\[-8pt]
\hline			     		       
         &                   &                  &              &              
         &                   &                  &                \\[-8pt]
$\tilde b_V$		     		       				    
         & $+1.57 (2)(1)$    & $+1.62(2)(2)$    &  N.A.        & $+1.106(11)$ 
         &  $+1.48(1)(1)$    &  N.A.            &  $+1.099(10)$  \\
$b_V$    & $+1.62 (3)(1)$    & $+1.68(3)(3)$    & $+1.54(2)$   & $+1.273(16)$ 
         &  $+1.54(1)(0)$    &  $+1.41(2)$      &  $+1.254(14)$  \\
$\tilde b_A-\tilde b_V$	     		       				    
         & $-0.56 (9)(8)$    & $-0.26(6)(9)$    &  N.A.        & $-0.002(0)$  
         &  $-0.11(4)(3)$    &  N.A.            &  $-0.002(0)$   \\
$\tilde b_P-\tilde b_S$	     		       				    
         & $-0.09 (9)(3)$    & $-0.02(6)(0)$    &  N.A.        & $-0.066(4)$  
         &  $-0.06(3)(1)$    &  N.A.            &  $-0.062(4)$   \\
$\tilde b_P-\tilde b_A$	     		       				    
         & $-0.20 (9)(7)$    & $-0.09(8)(6)$    &  N.A.        & $+0.002(4)$  
         &  $-0.07(2)(3)$    &  N.A.            &  $+0.002(0)$   \\
$\ds{\tilde b_A-\tilde b_P\atop{} +\tilde b_S/2}$			    
         & $+0.48 (1)(2)$    & $+0.69 (25)(1)$  &  N.A.        & $+0.584(7)$  
         &  $+0.50(5)(0)$    &  N.A.            &  $+0.579(6)$   \\
% $\ds{b_A-b_P\atop{}+b_S/2} $		       				    
%        & $+0.49 (2)(2)$    & $+0.72 (26)(1)$  &  N.A.        & $+0.673(9)$  
%        &  $+0.52(5)(0)$    &  N.A.            &  $+0.660(8)$   \\
         &                   &                  &              &              
         &                   &                  &                \\[-8pt]
\hline			     		       
         &                   &                  &              &              
         &                   &                  &                \\[-8pt]
$\tilde b_A$		     		       				    
         & $+1.01 (09)(09)$  & $+1.36(06)(10)$  &  N.A.        & $+1.104(11)$ 
         &  $+1.37(4)(2)$    &  N.A.            &  $+1.097(9)$   \\
$\tilde b_P$		     		       				    
         & $+0.81 (14)(16)$  & $+1.28(09)(15)$  &  N.A.        & $+1.105(11)$ 
         &  $+1.30(4)(5)$    &  N.A.            &  $+1.099(10)$  \\
$\tilde b_S$		     		       				    
         & $+0.90 (17)(13)$  & $+1.29(10)(12)$  &  N.A.        & $+1.172(30)$ 
         &  $+1.36(5)(4)$    &  N.A.            &  $+1.161(26)$  
			     		       		         \\[3pt]
\hline
\end{tabular}
\end{table*}
The results of our calculation done at \(\beta = 6.0\) and \(6.2\) are
presented in Table~\ref{tb:1} and compared to the previous estimates
by the ALPHA collaboration~\cite{alpha} and perturbation theory.  In
those cases where the coefficients \(b_X\) and \(\tilde b_X\) differ
by more than the estimated error in our determination, we quote both
for easier comparison with previous results.  

The \(\beta=6.0\) calculations were done on \(16^3 \times 48\)
lattices, with a sample of 83 configurations at \(c_{SW} = 1.4755\),
and 125 configurations at \(c_{SW}=1.769\).  At \(\beta=6.2\), we used
61 configurations with a lattice size of \(24^3 \times 64\).  The
region of axial rotation used for the Ward identities comprise the
time-slices \(4-18\) at \(\beta=6.0\) and \(6-25\) at \(\beta=6.2\).
Details of the quark masses and the lattice discretizations employed
in the calculation will be presented elsewhere~\inlinecite{lanlnext}.

\section{Dependence on \(\beta\) and \(c_{SW}\)}
The determination of \(c_A\) is illustrative of the improvement we
notice as we go to weaker couplings.  This quantity is
determined by requiring that the ratio 
\begin{equation}
\frac{ \sum_{\vec{x}} \langle 
 \partial_\mu [A_\mu + 
  a c_A \partial_\mu P]^{(ij)}(\vec{x},t) J^{(ji)}(0) \rangle} 
 {\sum_{\vec{x}} \langle P^{(ij)}(\vec{x},t) J^{(ji)}(0) \rangle} 
\end{equation}
be independent of the time \(t\) at which it is evaluated, up to errors
of \(O(a^2)\).  Because this flatness criterion is automatically
satisfied if the correlators are saturated by a single state, the
determination is very sensitive to the small \(t\) region.  On the
other hand, at very small \(t\), the \(O(a^2)\) errors dominate.  We
find that the region well fit by a constant starts at smaller values
of \(t\) at \(\beta=6.2\), and that this results in smaller
statistical errors on the determination of \(c_A\).  As \(c_A\) feeds
into the 3-pt Ward identity calculation, the statistical errors are
typically smaller at the weaker coupling.  In addition, the
sensitivity to how the continuum derivatives are discretized on the
lattice is reduced, leading to smaller systematic errors.

In contrast, the proper choice of \(c_{SW}\) at \(\beta=6.0\) has a
much smaller effect on the quality of the signal.  We find that
changing \(c_{SW}\) from 1.4755 to 1.769 has little effect on the
statistical and systematic errors but brings \(Z_X^0\) and \(c_X\)
closer to their perturbative values.  In addition, the differences
between the various \(\tilde b_X\), which are almost zero in
perturbation theory, also decrease.

\section{Comparison with previous results}
Different calculations of the non-perturbative renormalization
constants are expected to differ because of residual \(O(a^2)\) errors
in the theory.  The magnitude of these effects are expected to be
\(O(\Lambda_{QCD}^2 a^2)\) in \(Z_X^0\) and \(O(\Lambda_{QCD} a)\) in
\(c_X\) and \(b_X\).  Numerically these are about 0.02 and 0.15
respectively at \(\beta=6.0\), and 0.01 and 0.1 respectively at
\(\beta=6.2\).  

The differences between our values and those determined by the ALPHA
collaboration are consistent with this qualitative expectation.  In
fact, the differences between the two determinations of \(Z_V^0\) are
0.017(1) and 0.008(1) at \(\beta=6.0\) and 6.2 respectively, exactly
as expected from an \(O(a^2)\) scaling. For \(Z_A^0\), the differences
of 0.020(17) and 0.009(13) at the two \(\beta\) values are similar in
magnitude, but have much larger errors. 

Of the coefficients \(b_X\), only \(b_V\) has been calculated by the
ALPHA collaboration.  The corresponding differences, 0.14(5) and
0.13(2), have large errors but are not inconsistent with the expected
\(O(a)\) scaling.

The situation is less clear for the coefficients \(c_X\).  The ALPHA
collaboration has computed only \(c_A\) and \(c_V\) and our results
are completely consistent with theirs at \(\beta=6.2\).  The
differences at \(\beta=6.0\) are much larger in comparison and
marginally significant.

\section{Comparison with perturbation theory}
The perturbative results quoted in the table are based on one-loop
tadpole improved perturbation theory and their errors are obtained by
squaring the one-loop term in this tadpole improved series.  We notice
that the non-perturbatively determined \(Z_V^0\) and \(Z_P^0/Z_S^0\)
are significantly lower than their perturbative values and all the
\(\tilde b_X\)'s are higher.  The coefficients \(c_X\) are small, and
except for \(c_A\), agree with perturbation theory within errors.

It is worth mentioning that the difference between the perturbative
and non-perturbative \(Z_V^0\) is about 0.046(1) at \(\beta=6.0\) and
0.037(0) at \(\beta=6.2\), where the errors are purely statistical. As
an \(O(a^2)\) contamination in our non-perturbative determination is
expected to change by a factor of two between these two calculations,
it cannot be a dominant contribution to the observed difference.
Furthermore, noting that the tadpole improved \(\alpha_s\) is about
0.13 and 0.12 in the two calculations, these differences are only
about 2.5 times \(\alpha_s^2\), not unreasonable in a slowly
converging perturbative series.  The pattern is similar for
\(Z_P^0/Z_S^0\), where the difference between the perturbative and
non-perturbative results is about 0.09(2) at \(\beta = 6.0\) and
0.07(2) at \(\beta=6.2\). Because of the larger errors, it is,
however, not possible to rule out \(O(a^2)\) artifacts in this case.

Except for \(\tilde b_V\), the non-perturbative values for \(\tilde
b_X\) are constant within errors as we move from \(\beta=6.0\) to
\(\beta=6.2\).  The leading non-perturbative errors in these
quantities are \(O(a)\), which should change by a factor of about 1.4
between these two couplings.  On the other hand, the perturbative
one-loop term accounts for only about a fourth of \(\tilde b_V - 1\)
and \(\tilde b_A - 1\).  Such a high estimate of \(\tilde b_X\) with a
correspondingly low value of \(Z_X^0\) could indicate a problem with
understanding the quark mass dependences in general: more study is
needed to clarify this issue.

\section{Conclusion}
We have demonstrated the feasibility of our method for determining the
scheme-independent renormalization constants of the quark bilinear
operators.  Differences from previous calculations are consistent with
the expected \(O(a^2)\) differences.  Perturbation theory, even when
tadpole improved, seems to have \(O({\rm few} \times \alpha_s^2)\)
residual errors for the chirally extrapolated renormalization
constants.  The improvement constants \(c_X\) are small, but possibly
somewhat larger than predicted by perturbation theory; a detailed
comparison is still not possible due to the large errors.  The mass
dependences of the renormalization constants, given by \(\tilde b_X\),
are much larger than perturbation theory predicts.

\end{document}